# Overcoming stress limitations in SiN nonlinear photonics via a bilayer waveguide

Karl J. McNulty[1], Shriddha Chaitanya[1], Swarnava Sanyal[2], Andres Gil-Molina[1], Mateus Corato-Zanarella[1], Yoshitomo Okawachi[2], Alexander L. Gaeta[1,2], and Michal Lipson[1,2]

[1]Department of Electrical Engineering, Columbia University, New York, New York, 10027, USA

[2]Department of Applied Physics & Applied Mathematics, Columbia University, New York, New York, 10027, USA

**Abstract:**

Silicon nitride (SiN) formed via low pressure chemical vapor deposition (LPCVD) is an ideal material platform for on-chip nonlinear photonics owing to its low propagation loss and competitive nonlinear index. Despite this, LPCVD SiN is restricted in its scalability due to the film stress when high thicknesses, required for nonlinear dispersion engineering, are deposited. This stress in turn leads to film cracking and makes integrating such films in silicon foundries challenging. To overcome this limitation, we propose a bilayer waveguide scheme comprised of a thin LPCVD SiN layer underneath a low-stress and low-index PECVD SiN layer. We show group velocity dispersion tuning at 1550nm without concern for film-cracking while enabling low loss resonators with intrinsic quality factors above 1 million. Finally, we demonstrate a locked, normal dispersion Kerr frequency comb with our bilayer waveguide resonators spanning 120nm in the c-band with an on-chip pump power of 350mW.

**Introduction:**

Low pressure chemical vapor deposition (LPCVD) silicon nitride (SiN) films have emerged as an ideal candidate for optical non-linear applications, owing to SiN's low optical loss and high $n_2$ value [1]. Despite the desirable characteristics, LPCVD SiN waveguides often require thick films (>400nm) to satisfy the group velocity dispersion (GVD) requirements for nonlinear processes in the telecom range. Such thick SiN films are plagued by high tensile stress and subsequent film cracking, thereby lowering the photonic device yield and limiting their applicability in large-scale systems. Figure 1a shows the simulated GVD for an LPCVD SiN waveguide with a 1.5um width and various heights. As an example of dispersion needed for nonlinear processes, we highlight the dispersion range that facilitates comb formation in the telecom wavelength range [2-11]. As Figure 1a indicates, the waveguide GVD in the C-band only reaches the target normal and anomalous GVD range when the waveguide thickness is beyond 500nm, above the typical thickness limit of 400nm in foundries [12]. Figure 1b shows an example of the extreme cracking in 700 nm SiN film which forms at the edge of a 100mm wafer (right) and propagates to the center of the wafer (left). Such cracking is almost unavoidable in LPCVD SiN films above the thickness threshold of 400nm and is a key limiting factor of wider integration of SiN based nonlinear photonics.

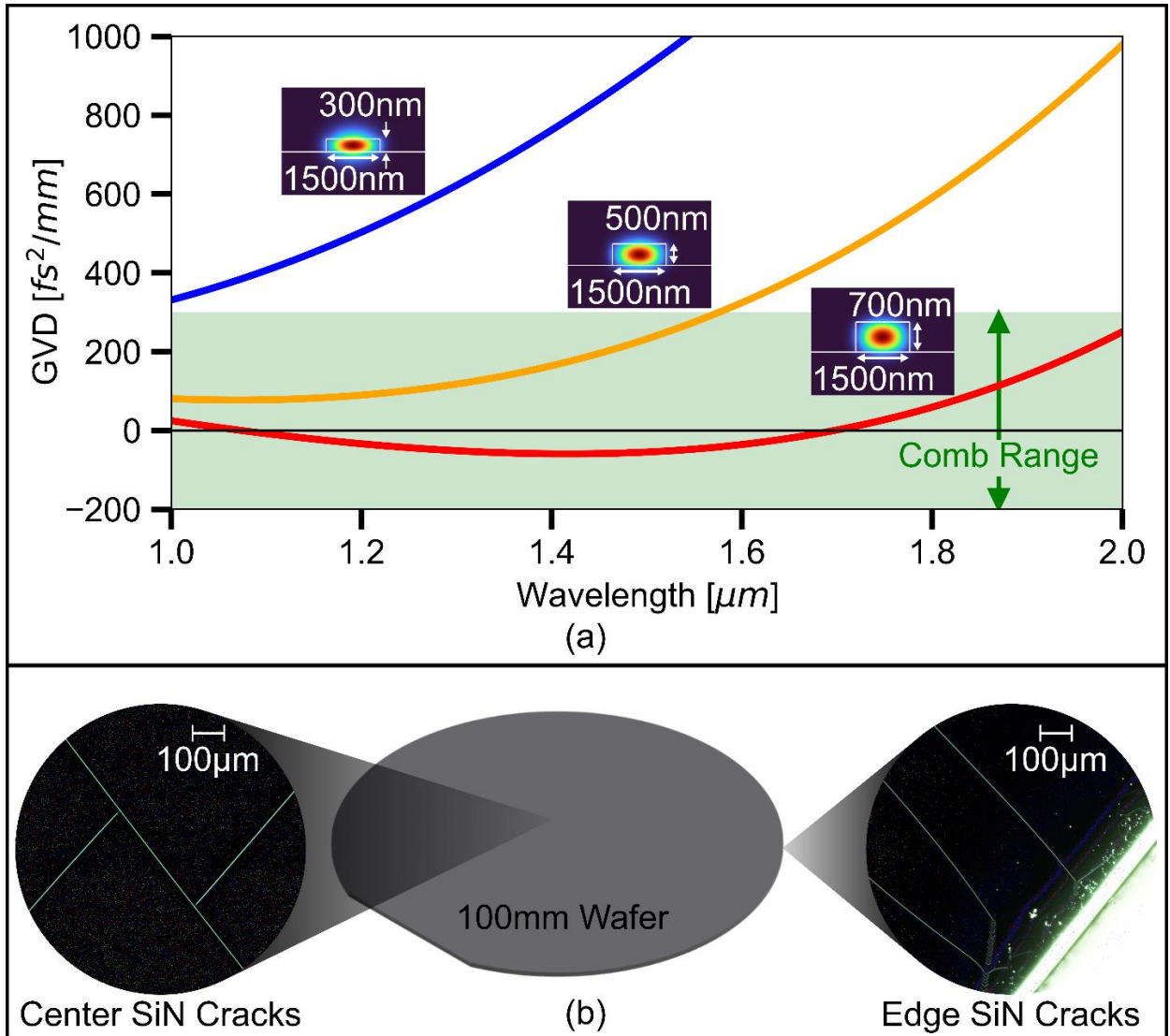

**Figure 1:** (a) Dispersion comparison of a 1.5um wide LPCVD SiN waveguide with various thicknesses. The green shaded area indicates a general GVD range which has been shown to generate microcombs. (b) Visible film cracking seen under a dark field microscope of a 730nm thick LPCVD SiN film on a 100mm wafer. Cracks propagate from the edge of a wafer (right) and pass through the center of the wafer (left).

To mitigate the material stress limitation, fabrication approaches consist of stress relief trenches utilized with and without thermal cycling techniques during the SiN deposition [13-18] or the use of chemical mechanical planarization (CMP) for a photonic damascene process [19-21] These processes suffer from film cracking at the relief trenches or dishing from different material removal rates during CMP. Such processes have also been limited to smaller wafer sizes, and crack-free, thick LPCVD SiN films have only been demonstrated up to a 150mm wafer scale

[14]. Circumventing LPCVD SiN through the use of sputtered SiN [22-24], PECVD SiN [25], and deuterated PECVD SiN [26-28] for stress-free, nonlinear devices have also been shown, but all have yet to reach either the low loss or consistency of LPCVD SiN for nonlinear applications.

Thin LPCVD SiN films below 400nm, in contrast to thick LPCVD SiN films, do not suffer from stress and are readily available in foundries. Nonlinear processes may be realized with such thin films via mode crossing techniques [29-31]. However, devices based on thinner nitride films lack high modal confinement and the ability for broadband dispersion engineering [32].

Here we endow a foundry compatible, thin film LPCVD SiN with the desired GVD via a stress-free bilayer waveguide configuration. Shown in Figure 2b, we deposit a low index PECVD SiN (index of 1.78 at 1550nm) on a thin film of LPCVD SiN. The low index SiN layer enables stress free tuning of the overall dispersion of the composite waveguide, while maintaining low loss for the whole composite waveguide. It also ensures high modal overlap with the bottom LPCVD SiN to utilize its higher nonlinearity as the low index SiN is less dense and, therefore, has a lower relative $n_2$ [33]. As an example of application, we use our bilayer waveguides to construct dual resonators, depicted in Figure 2a, which are commonly used for comb generation [10]. We choose a waveguide width of 1700nm for our design, an LPCVD SiN layer of 340nm in thickness, and a low index SiN layer of 400nm in thickness. As Figure 2c shows, our design allows significant control over the waveguide GVD compared to the unaided thin LPCVD SiN waveguide of the same width. From dispersion simulations, the bilayer waveguide exhibits a GVD within the c-band sufficiently low to where normal GVD SiN microcombs have been demonstrated [2, 4-7, 10]. In contrast we see that the 340nm thick LPCVD SiN waveguide exhibits a very high normal GVD and fails to reach the generally required dispersion for microcomb generation.

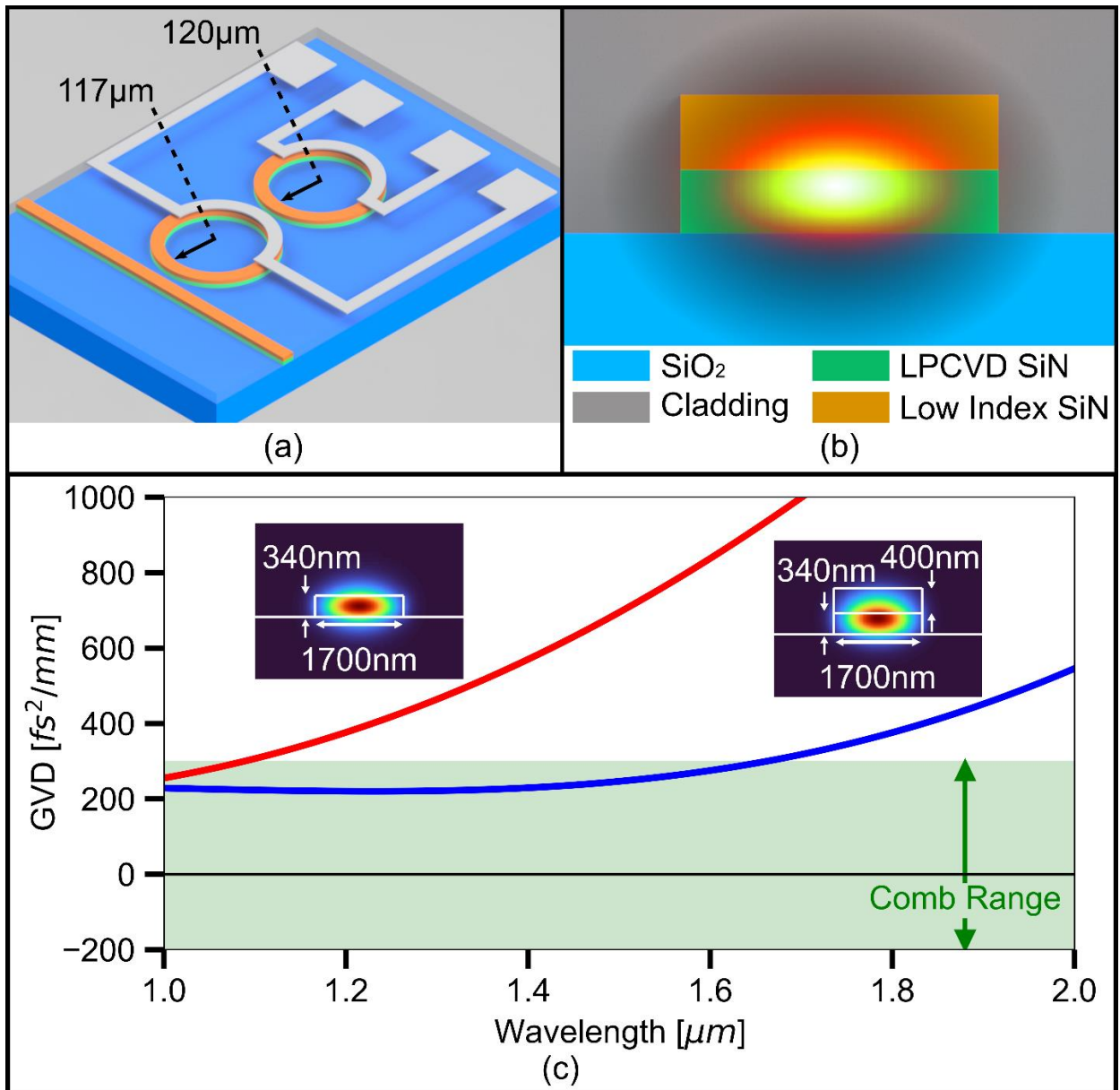

**Figure 2:** (a) 3D rendering of the bilayer design consisting of dual ring resonators and on-chip heaters for resonance detuning control. (b) Cross-section of the 1700nm x 740nm waveguide along with an overlay of the optical mode. (c) Waveguide GVD comparison of an unaided 1700nm x 340nm LPCVD SiN waveguide with that of the bilayer design. The addition of the low index SiN layer brings the dispersion down within the range of demonstrated combs without the tradeoff of high film stress.

**Results and Discussion:**

We experimentally validate that our waveguide design enables low loss resonators comparable to the loss levels of single-core LPCVD SiN resonators. To directly compare device loss, we fabricate identical single ring resonators with (a) standard 730nm thick LPCVD SiN waveguides,

(b) our bilayer waveguides with the thicknesses specified earlier, and (c) 730nm thick low index SiN only waveguides. We choose the ring radius to be 150um and the waveguide width to be 2000nm for all devices tested. To minimize difference in performance attributed to process variation, these waveguides receive the same process steps and are fabricated in parallel with each other. For loss characterization, we selected one undercoupled ring for each waveguide type and measured roughly 50 resonances across a spectrum of 1500nm-1600nm. We then employ a coupled mode model with backscattering [34] to fit each individual resonance and extract the intrinsic quality factor as a measurement of cavity loss. Figure 3 shows a histogram of the fitted intrinsic quality factors for each device spectrum. Despite a slight reduction in quality factor compared to LPCVD SiN, one can see that our bilayer waveguides retain intrinsic quality factors well above 1 million across the c-band. This is in stark comparison to the single-core low index SiN waveguide which has much lower intrinsic quality factors in the range of 100-500k. Thus, our dispersion tuning design maintains low optical loss while not suffering from film stress.

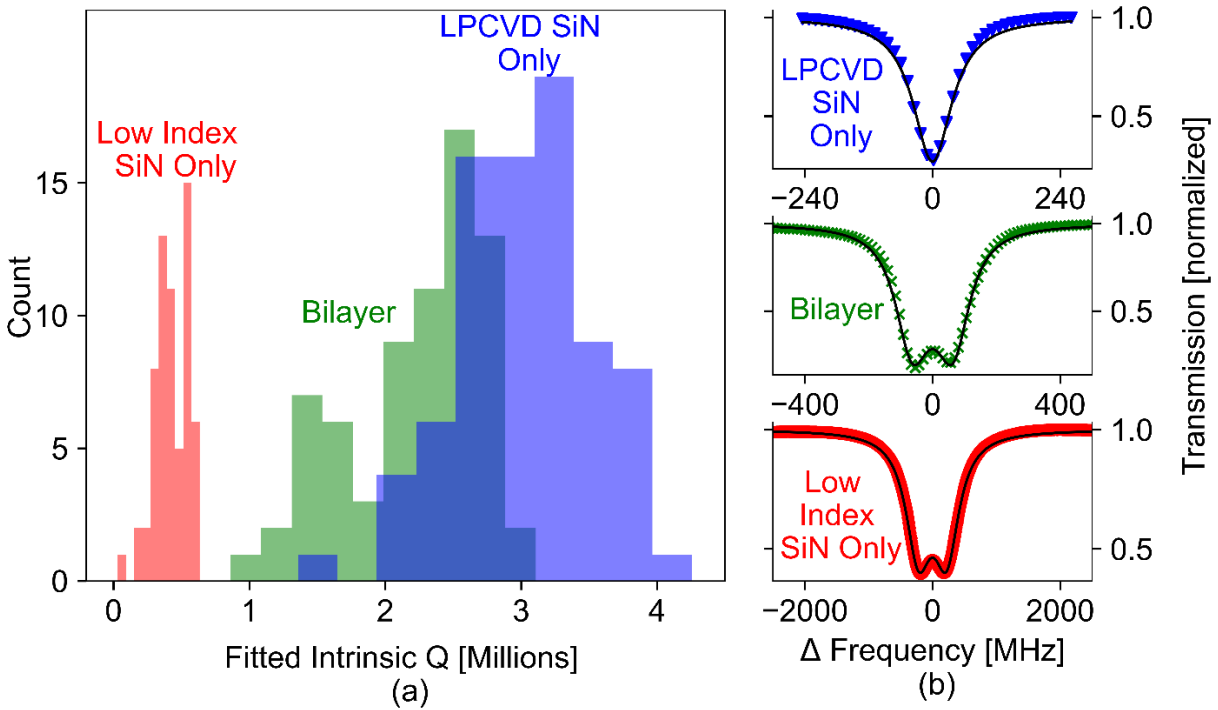

**Figure 3:** (a) Histogram of the fitted intrinsic quality factors measured across a spectrum of 1500nm-1600nm for three separate ring resonator devices. All three rings were 150um in radius and had cross-sections of 2000nm x 730nm. (b) Sample resonance fits for each of the different rings.

To demonstrate the practicality of our bilayer approach for nonlinear photonic applications, we demonstrate a low-noise, Kerr frequency comb spanning approximately 120nm with 350mW of on-chip pump power. As our bilayer waveguide has a GVD shown in Figure 2b, we employ the dual ring configuration, as mentioned above, to enable comb generation in the normal dispersion regime. For our comb device, we choose a width of 1700nm for our bilayer waveguide with a main ring of radius 120um and an auxiliary ring of radius 117um (as shown in Figure 2a). Pumping around 1557nm and following [10], we utilize our fully integrated, on-chip microheaters to tune the main ring and auxiliary ring frequency detunings to navigate to a low noise comb state. Figure 4 shows both the unlocked state (top) prior to tuning the microheater currents, and the low-noise, locked state (bottom) of the comb after tuning the microheater currents. Once locked, we observe the characteristic low RF frequency noise associated with such a locked state [35], in stark contrast to the high RF frequency noise seen in the unlocked state.

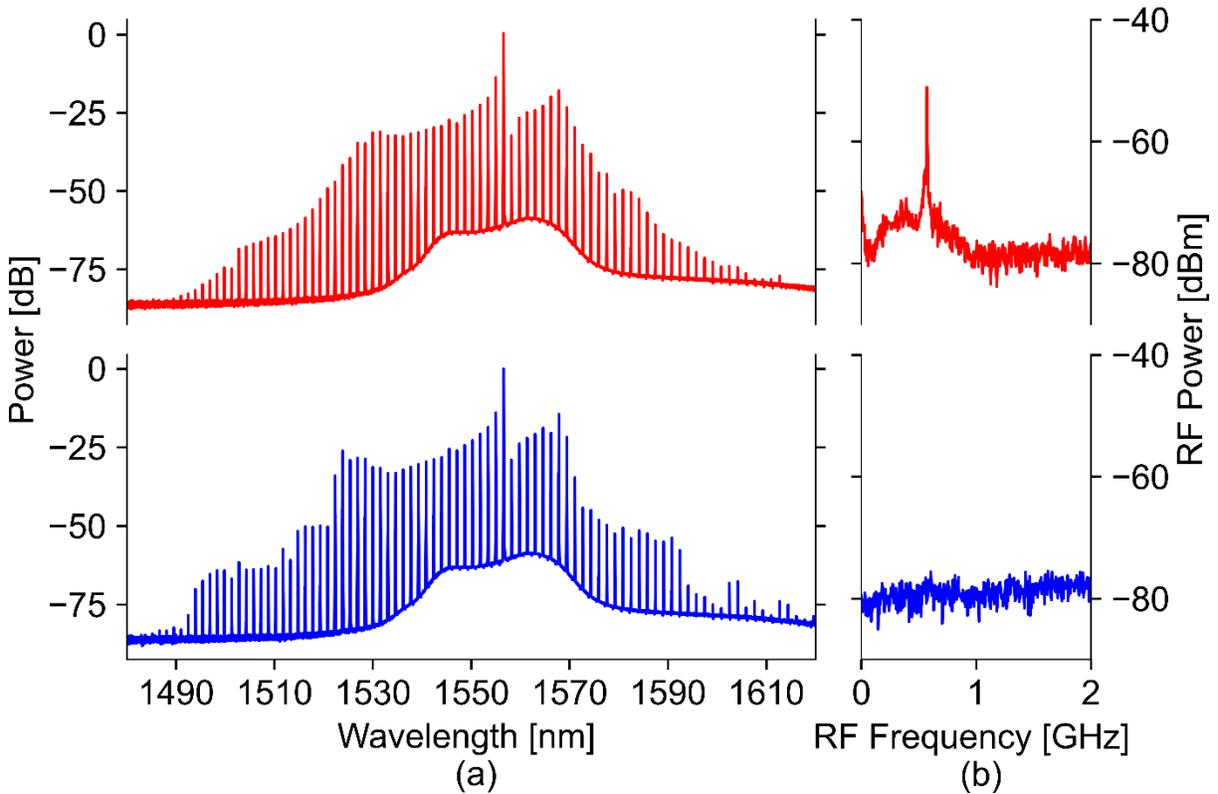

**Figure 4:** (a) OSA spectrum of the unlocked (top) and locked (bottom) comb states of our bilayer waveguide resonator (normalized to the pump line). (b) Associated RF frequency noise measured via an electronic spectrum analyzer (ESA).

**Conclusion:**

Our approach to dispersion engineering provides a solution to the film stress plaguing SiN technology. By decoupling the waveguide dispersion from the thickness of a single LPCVD SiN layer, we enable GVD dependent nonlinear processes without compromising on optical loss or film stress during fabrication. We generate coherent, low-noise combs using our new waveguide design and show that our approach is compatible with mode-crossing techniques previously used for microcomb generation. Moreover, this approach is likely adaptable to other materials, as other PECVD $Si_xN_y$ films or possibly $TiO_2$ could be utilized instead.

**Fabrication Methods:**

We fabricate our devices starting from 100mm wafers with 4um of thermal oxide. We then deposit the 340nm of bottom SiN via LPCVD and follow this by depositing the 400nm layer of low index SiN via PECVD. We utilize an oxide hardmask alongside electron beam lithography to define our waveguides and etch them with a plasma dry etch. The waveguides are then annealed in an argon environment for 3 hours and subsequently clad with a high temperature silicon dioxide. The platinum heaters are defined on the cladding oxide through a lift off process, and we perform a final deep silicon etch prior to dicing to enable low loss edge couplers.

**Acknowledgements:**

The photonic chip fabrication was done in part at the Columbia Nano Initiative (CNI) Shared Lab Facilities at Columbia University, and in part at the Cornell NanoScale Facility, a member of the National Nanotechnology Coordinated Infrastructure (NNCI), which is supported by the National Science Foundation (Grant NNCI-2025233). The authors also acknowledge the use of the Nanofabrication Facility at the Advanced Science Research Center at The Graduate Center of City University of New York.

**Funding:**

The authors acknowledge the financial support of the Air Force Office of Scientific Research under grant FA9550-20-1-0297 and the Defense Advanced Research Projects Agency (DARPA) under grant HR00112420368.

**Author Contributions:**

KJM, YO, ALG, and ML conceived the idea. KJM did the planning and execution of the experiment including design, simulations, fabrication, testing, and data analysis. SC aided in film deposition. SS, AGM, and MCZ aided in device testing. KJM and ML wrote the manuscript. Edits were provided by all authors. All authors have accepted responsibility for the entire content of this manuscript and approved its submission.

**Conflict of Interest:**

Authors state no conflict of interest

**Data Availability:**

Data underlying the results are presented throughout the paper and additional data may be obtained from the authors upon reasonable request.